\documentclass[prb,twocolumn,showpacs]{revtex4}
\usepackage{amsmath}
\usepackage{amssymb}
\usepackage{dcolumn}
\usepackage{graphicx}

\begin{document}

\title[Short title for running header]{What does the giant thermal Hall effect observed in the high temperature superconductors imply?}
\author{Tao Li}
\affiliation{Department of Physics, Renmin
University of China, Beijing 100872, P.R.China}
\date{\today}

\begin{abstract}
A giant negative thermal Hall signal is discovered recently in the pseudogap phase of the high temperature superconductors\cite{Taillefer}. The Wiedemann-Franz law relating the charge and thermal transport is found to be strongly violated as the thermal Hall signal increases monotonically with decreasing doping in the pseudogap phase when the Hall response of the system is seen to be strongly suppressed. In particular, the thermal Hall signal is the strongest in the parent compounds, which are antiferromagnetic insulators. People believe that such an observation may challenge our understanding of the parent compounds as a spin-$\frac{1}{2}$ Heisenberg antiferromagnet on the square lattice, since it is forbidden by a well known no-go "theorem" for quantum magnets with an edge-sharing lattice. Here we show that the observed thermal Hall signal can be naturally understood as the orbital magnetic response of the quantum Heisenberg model on square lattice. We show that the thermal Hall signal is in fact a measure of the spin chirality fluctuation in the system. The universal observation of the giant negative thermal Hall signal in and only in the pseudogap phase implies that the antiferromagnetic correlation between local spins is at the root of the pseudogap phenomena. It also implies that the pseudogap phenomena in the cuprates is always accompanied by strong spin chirality fluctuation, which is suppressed only when the local spin fluctuation at low energy is totally replaced by itinerant quasiparticle behavior at sufficiently large doping. 
\end{abstract}

\maketitle
The origin of the pseudogap phenomena in the high temperature superconductors remains elusive. There are many clues implying the importance of the antiferromagnetic correlation between the local spins for the development of such a phenomena. For example, ARPES measurements find that the pseudogap opens right at the temperature when the $^{63}Cu$ NMR spin relaxation rate, a measure of the antiferromagnetic spin fluctuation in the system, reaches its maximum\cite{Hashimoto}. More recently, thermodynamic measurement finds that the pseudogap phenomena ends suddenly at a quantum critical point around $x_{c}\approx0.2$, when the Fermi surface changes its topology from hole-like to electron-like\cite{Michon}. Although the origin of such a critical behavior is still elusive, its observation is consistent with a picture in which antiferromagnetic short-range correlated local spins are coupled to an itinerant quasiparticle system with coincident Van Hove singularity and antiferromagnetic hot spot\cite{Li}.    

Very recently, a giant negative thermal Hall signal is found universally in the high temperature superconductors just below the critical doping for pseudogap phase\cite{Taillefer}. The Wiedemann-Franz law relating the charge and thermal transport is found to be strongly violated. More specifically, the thermal Hall signal in the pseudogap phase is found to increase monotonically with decreasing doping, when the Hall response of the system is seen to be strongly suppressed. In particular, the thermal Hall signal becomes the strongest in the half-filled parent compounds, which are antiferromagnetic insulators. These observations indicate that the observed thermal Hall signal should be attributed to charge neutral degree of freedom that exist already in the parent compounds, most likely the local spins\cite{phonon}.  An important characteristic of the observed thermal Hall signal is that it increases linearly with the applied magnetic field at weak field, which implies that it is a probe of some property of the zero field phase.

The local spin in the parent compounds of the high temperature superconductors is known to be well described by the spin-$\frac{1}{2}$ Heisenberg antiferromagnetic model on the square lattice. These local spins and their antiferromagnetic correlation remain robust even in strongly doped systems, as implied by the magnon-like dispersive mode revealed by recent RIXS measurements\cite{RIXS}. However, previous studies suggest that most mechanisms proposed for the thermal Hall effect are ineffective on the square lattice as a result of a no-go theorem: the chiral contributions from neighboring unit cell cancel with each other on edge-sharing lattice, in either the linear spin wave theory or the RVB mean field theory treatment\cite{Katsura,Hotta,PALee,Sachdev}. This leads some researchers to propose that the observation of the giant thermal Hall signal in the cuprates may imply a failure of the spin-$\frac{1}{2}$ Heisenberg antiferromagnetic model on the square lattice to describe the physics of their parent compounds\cite{Sachdev,Xu,ZXLi}.

\begin{figure}
\includegraphics[width=8cm]{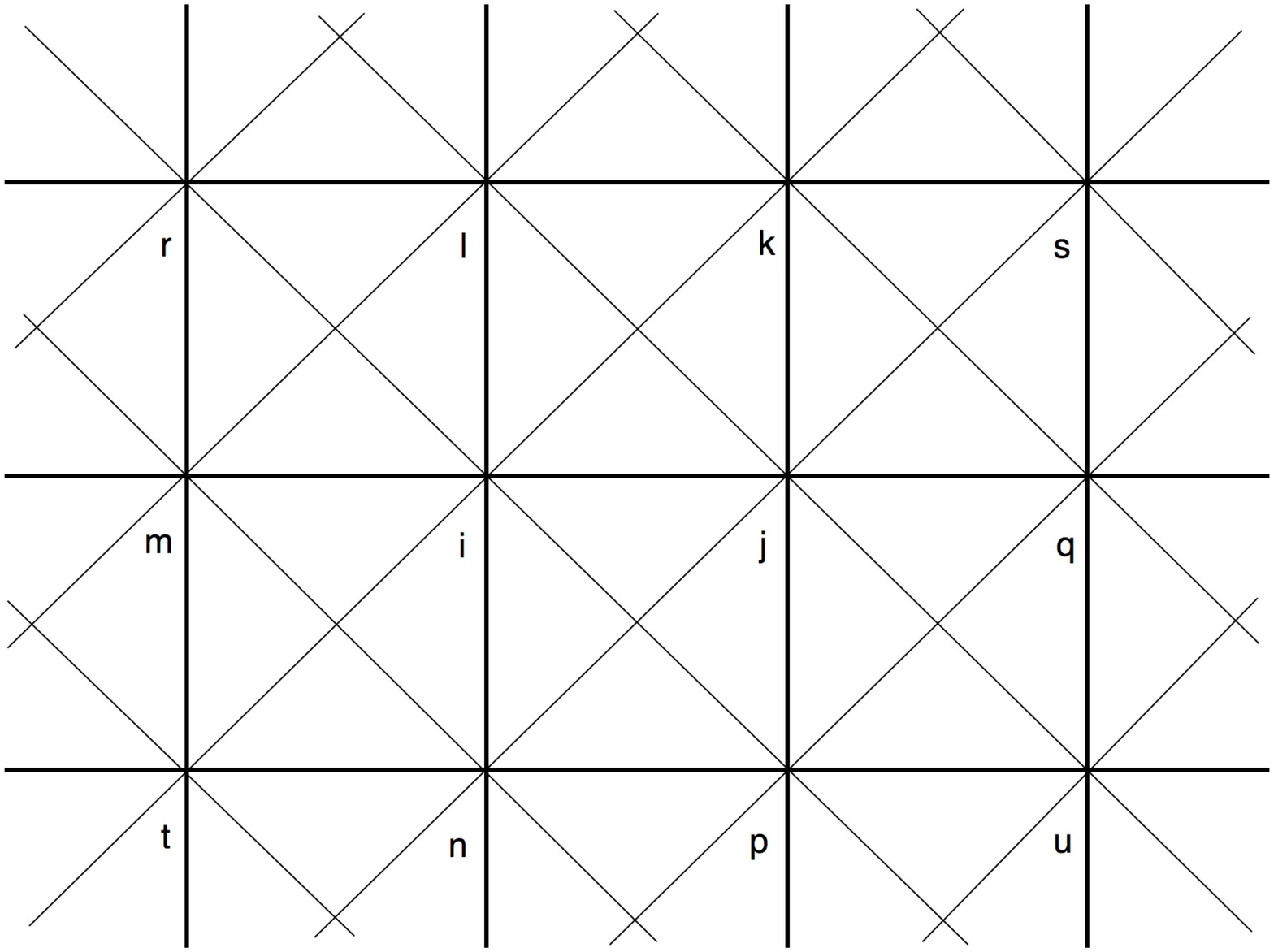}
\caption{Illustration of the model studied in this work. The Heisenberg part of the model, $H_{J}$, is the sum of antiferromagnetic exchange interaction on nearest bonds(denoted here as thick bonds). The ring-exchange part of the model, $H_{3}$, consists of the sum of orbital coupling of the magnetic field with the scalar spin chirality on all the elementary triangles of the lattice. An elementary triangle(for example, the triangle $\bigtriangleup_{i,j,k}$) is made of two nearest bonds and one next-nearest bonds(denoted as thin bonds). Note that there are four elementary triangles in each unit cell. According to strong coupling expansion of the Hubbard model, the ring-exchange coupling constant is given by $J_{3}=-\frac{24t^{2}t'}{U^{2}}$, in which  $t$ and $t'$ are the hopping integrals on the thick and thin bonds. The site indices are introduced for later reference.}
\end{figure}

Here we show that the observed giant negative thermal Hall signal can be naturally understood as the chiral response of the spin-$\frac{1}{2}$ antiferromagnetic Heisenberg model on the square lattice with a sizable multi-spin exchange coupling. The Heisenberg part of the model is given by
\begin{equation}
H_{J}=J\sum_{<i,j>} \mathrm{S}_{i} \cdot \mathrm{S}_{j}.
\end{equation}
Here the sum is over nearest neighboring sites. We define $|\mathrm{HAF}\rangle$ as the ground state of $H_{J}$ for latter convenience. The external magnetic field can couple to the local spin through either the Zeeman term or the multi-spin exchange process. Here we will neglect the Zeeman coupling\cite{DM}. Up to the third order in the strong coupling expansion of the Hubbard model, we have the following three spin ring-exchange term
\begin{equation}
H_{3}=-J_{3}\sin(\Phi_{3}) \sum_{\bigtriangleup_{i,j,k}} \ \mathrm{S}_{i}\cdot (\mathrm{S}_{j}\times\mathrm{S}_{k})\nonumber.
\end{equation} 
Here $J_{3}=-\frac{24t^{2}t'}{U^{2}}$ is the three spin ring-exchange coupling constant around an elementary triangle $\bigtriangleup_{i,j,k}$ on the square lattice\cite{ring}, in which $t$ and $t'$ are the nearest and next-nearest neighboring hopping integrals. The additional minus sign in front of $J_{3}$, which can be easily overlooked but is crucial for the following discussion, comes from the fact that the electron feel the magnetic flux with a negative charge. Site $i,j,k$ in $\bigtriangleup_{i,j,k}$ are always ordered in the anti-clockwise manner(see Fig.1). $\Phi_{3}$ is the magnetic flux enclosed in $\bigtriangleup_{i,j,k}$. \ $\hat{\chi}_{i,j,k}=\mathrm{S}_{i}\cdot (\mathrm{S}_{j}\times\mathrm{S}_{k})$ is the scalar spin chirality in the triangle $\bigtriangleup_{i,j,k}$. For later convenience, we define $\hat{\mathcal{C}}= \sum_{\bigtriangleup_{i,j,k}} \hat{\chi}_{i,j,k}$ as the total spin chirality of the system. According to experimental fit for the parent compound La$_{2}$CuO$_{4}$, $t'\approx-0.4t$, $U\approx7t$, so that $J_{3}\approx 0.3 J$, which is not at all small\cite{LCO}.

$H_{3}$ has the right symmetry to generate a nonzero thermal Hall signal since it is odd under both time reversal and mirror reflection. However, it is generally believed that on the square lattice $H_{3}$ can not play any role at weak field. More specifically, it can be shown that its contribution vanishes identically at weak field in either the linear spin wave theory or the RVB theory at the mean field level\cite{Katsura,Hotta,PALee,Sachdev}. In the linear spin wave theory, such a null result is easily expected from the collinear ordering pattern in the ground state, which is incompatible with a nonzero spin chirality. In fact, if we approximate the local spins as classical vectors of length $\frac{1}{2}$, one can prove that the classical ground state of the model is totally unchanged by $H_{3}$ for $J_{3}\sin\Phi_{3}<\frac{J}{2}$(see Appendix A). 

The situation in the RVB theory is more involved. Here we will adopt the Bosonic RVB theory\cite{Fermion}, in which the spin operator is written as $\mathrm{S}_{i}=\frac{1}{2}\sum_{\alpha,\beta}b^{\dagger}_{i,\alpha}\sigma_{\alpha,\beta}b_{i,\beta}$. $b_{i,\alpha}$ is a Schwinger Boson operator with spin $\alpha$. In the Schwinger Boson formulation, we can define two spin rotationally invariant objects between a pair of sites, namely, the paring field $\hat{A}_{i,j}=\frac{1}{\sqrt{2}}\sum_{\alpha}\ \alpha \ b_{i,\alpha}b_{j,-\alpha}$ describing the antiferromagnetic correlation and the hopping field $\hat{B}_{i,j}=\frac{1}{\sqrt{2}}\sum_{\alpha}\  \ b^{\dagger}_{i,\alpha}b_{j,\alpha}$ describing the ferromagnetic correlation. The Heisenberg part of the model can be written in terms of the pairing field as
\begin{equation}
H_{J}=-\frac{J}{2}\sum_{<i,j>}\hat{A}^{\dagger}_{i,j}\hat{A}_{i,j}+C,
\end{equation}
In the mean field treatment, the ground state of $H_{J}$ is known to be described by the following mean field ansatz
\begin{eqnarray}
A_{i,i+x}=A_{i,i+y}=A\nonumber\\
B_{i,i+x}=B_{i,i+y}=0,
\end{eqnarray}
in which $A_{i,j}=\langle \hat{A}_{i,j} \rangle$, $B_{i,j}=\langle \hat{B}_{i,j} \rangle$ are the expectation value of the pairing and hopping field in the mean field ground state \cite{LDA}. We will show that within the mean field treatment, it is impossible to generate a linear-in-field thermal Hall response around this saddle point when we turn on $H_{3}$.  

In the Schwinger Boson formulation, the spin chirality operator is given by\cite{Messio} 
\begin{equation}
\mathrm{S}_{i}\cdot (\mathrm{S}_{j}\times\mathrm{S}_{k})=\frac{i}{\sqrt{2}}(\hat{B}_{i,j}\hat{B}_{j,k}\hat{B}_{k,i}-h.c.).
\end{equation}
If we approximate $\hat{B}_{i,j}$ with their mean field values, the contribution of $H_{3}$ will vanish at the linear order since the expectation value of $\hat{B}_{i,j}$ between nearest neighboring sites is zero when the ansatz Eq. (3) is assumed. More generally, we can decouple $H_{3}$ in both the hopping and the pairing channel. However, one find that the contribution from $H_{3}$ still vanishes at the linear order(see Appendix B). In fact, one can show with numerical optimization that the expectation value of $H_{3}$ is always zero in the Schwinger Boson mean field theory for even rather large value of $J_{3}\sin\Phi_{3}$, no matter how we modify the form of the mean field ansatz(see Appendix C).

However, it is obvious that these null results are just artifacts of the semiclassical or the mean field approximation adopted. More specifically, while the expectation value of the spin chirality is zero in $|\mathrm{HAF}\rangle$, its fluctuation in $|\mathrm{HAF}\rangle$ is nonzero. $H_{3}$ can thus generate virtual excitation on $|\mathrm{HAF}\rangle$ and induce a nonzero spin chirality. At weak field, the induced scalar spin chirality can be estimated from perturbation theory and is given by 
\begin{equation}
\mathcal{C}=\langle\ \hat{\mathcal{C}}\ \rangle\simeq -J_{3}\sin{\Phi_{3}}\sum_{n\neq0}\frac{|\langle n |\ \hat{\mathcal{C}}\ |\mathrm{HAF}\rangle|^{2}}{E_{0}-E_{n}}.
\end{equation}
Here $| n \rangle$ and $E_{n}$ denotes the eigenstates and the corresponding eigenvalues of $H_{J}$. As will be clear below, the thermal Hall signal is determined by the derivative of $\mathcal{C}$ over temeprature, rather than its absolute value. The temperature dependence of the field induced scalar spin chirality is given by
\begin{equation}
\mathcal{C}(T) \simeq -J_{3}\sin{\Phi_{3}} \frac{1}{Z}\sum_{n,m}\frac{e^{-\beta E_{m}}-e^{-\beta E_{n}}}{E_{n}-E_{m}}|\langle n | \hat{\mathcal{C}} |m \rangle |^{2}
\end{equation}

Now we show that the thermal Hall conductivity of the system is given by the temperature dependence of $\mathcal{C}(T)$ as $\kappa_{xy}(T)=-\frac{J^{2}}{2}\frac{\partial \chi(T)}{\partial T}$, in which $\chi(T)=\frac{\mathcal{C}}{4N}$ is the scalar spin chirality per triangle and $N$ is the number of unit cell. The key point is to note that the energy current operator of the Heisenberg model can be expressed directly in terms of the spin chirality operators\cite{Shastry,Han}. More specifically, if we define $H_{i}=\frac{J}{2}\sum_{\delta}\mathrm{S}_{i}\cdot\mathrm{S}_{i+\delta}$ as the local energy of the Heisenberg model at site $i$(with site $i+\delta$ as its four nearest neighbors), the energy current from site $i$ to site $j$ is given by
\begin{eqnarray}
J^{E}_{i,j}&=&\frac{1}{i\hbar}[H_{j},H_{i}]\nonumber\\
=\frac{J^{2}}{4}&[&(\hat{\chi}_{i,j,l}+\hat{\chi}_{i,j,k}-\hat{\chi}_{i,n,j}-\hat{\chi}_{i,p,j})\nonumber\\
&+&(\hat{\chi}_{m,i,j}+\hat{\chi}_{q,i,j})\ ]
\end{eqnarray}
Here the meaning of the site indices $i,j,k,l,m,n,p,q$ can be found in Fig. 1. $H_{3}$ will also contribute to the energy current at the linear order of $J_{3}\sin\Phi_{3}$. As is shown in Appendix D, this contribution, denoted as $J^{E(1)}_{i,j}$ there, measures the fluctuation of vector spin chirality in the system. One find that following analysis on $J^{E}_{i,j}$ also applies for $J^{E(1)}_{i,j}$. We will thus focus on $J^{E}_{i,j}$ for the moment.

In the bulk of the system at equilibrium, the expectation value of $J^{E}_{i,j}$ is exactly zero as a result of the PT symmetry. More specifically, $J^{E}_{i,j}$ is odd under time reversal but even with respect to the reflection about a line through site $i$ and site $j$. For example, the expectation value of $\hat{\chi}_{i,j,l}+\hat{\chi}_{i,j,k}-\hat{\chi}_{i,n,j}-\hat{\chi}_{i,p,j}$ vanishes identically when the system establishes a spatially uniform scalar spin chirality. At the same time, the expectation value of $\hat{\chi}_{m,i,j}+\hat{\chi}_{q,i,j}$ is zero as a result of the PT symmetry. 

The situation becomes totally different at the boundary of the system, where the reflection symmetry is broken. As illustrated in Fig. 2, on the upper edge of the system(suppose that it is along the x-direction), the spin chirality in the triangles inside the boundary will no longer be compensated by that outside the boundary. We thus expect a nonzero edge energy current of the size
\begin{equation}
J^{E}\approx-\frac{J^{2}}{2}\chi,\nonumber
\end{equation}
in which $\chi$ is the expectation value of spin chirality in a triangle\cite{relax}. From the PT symmetry of the system, we expect a counter-propagating energy current at the lower edge of the system with exactly the same magnitude. 

\begin{figure}
\includegraphics[width=8.5cm]{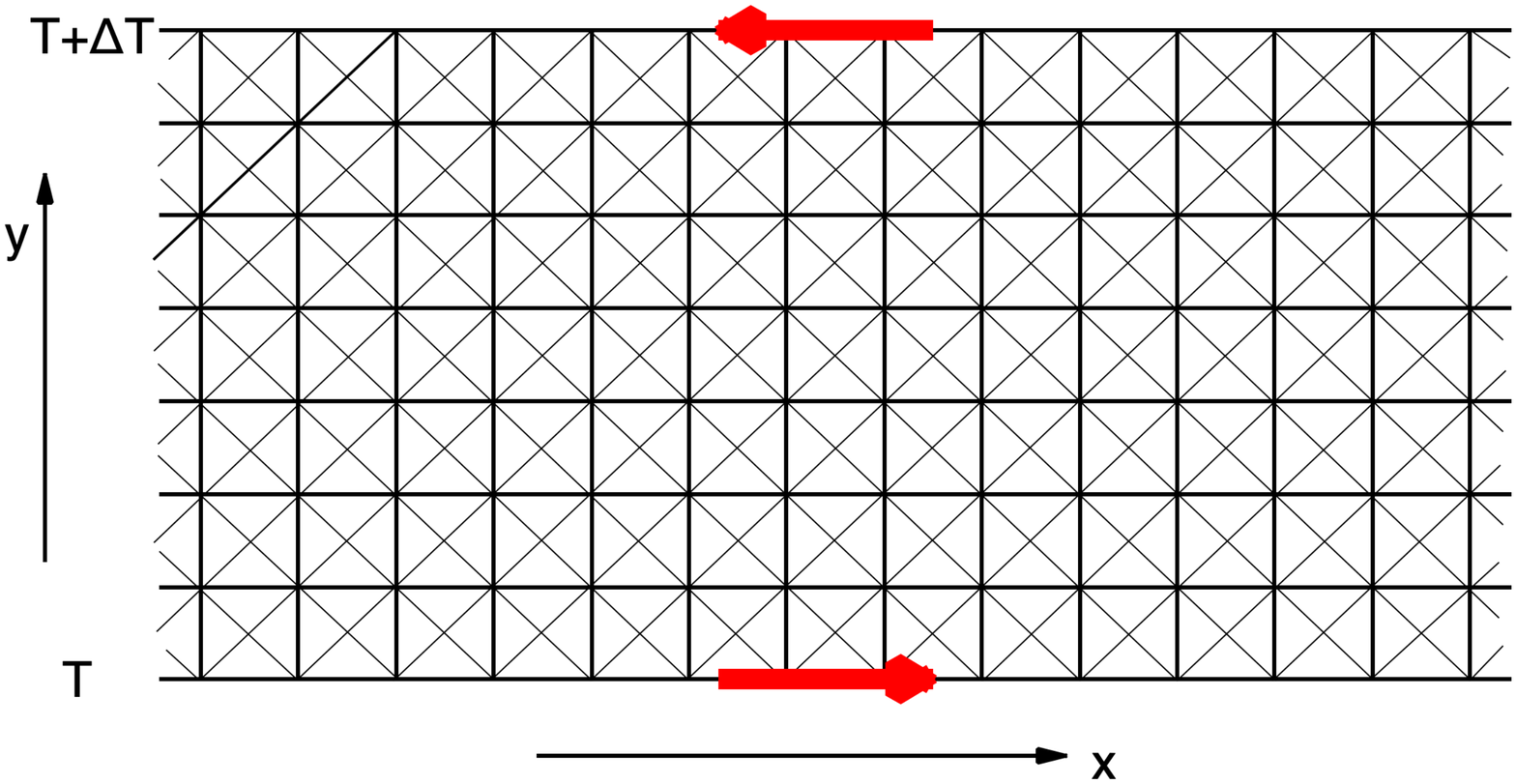}
\caption{Emergence of the edge energy current in system with a uniform spin chirality. When a temperature gradient is applied along the y-direction, the balance between the energy current at the upper and the lower edge is broken and a transport energy current along the x-direction is generated.}
\end{figure}

When the system is at its equilibrium, the energy current on the upper edge is compensated by that on the lower edge and the total transport energy current in the x-direction is zero. The application of a transverse temperature gradient in the y-direction will break such a balance and induce a net transport energy current of the size\cite{leq} 
\begin{equation}
J^{\mathrm{tr}}\approx -\frac{J^{2}}{2}[ \chi(T+\Delta T)-\chi(T)]\approx \kappa_{xy}(T)\Delta T,
\end{equation}
in which 
\begin{equation}
\kappa_{xy}(T)=-\frac{J^{2}}{2} \frac{\partial \chi(T)}{\partial T}. 
\end{equation}
The thermal Hall effect thus measures the temperature dependence of the field induced spin chirality in the system and is thus naturally linear in the applied magnetic field at weak field. 

Let us first determine the sign of the thermal Hall signal. Since $t'/t<0$ in cuprates, $J_{3}>0$. We thus have $\chi>0$ for magnetic field applied in the $+z$ direction. The circulating energy current is thus running in the $+x$($-x$) direction at the lower(upper) edge. Since the field induced spin chirality decreases with increasing temperature, the transport energy current should flow in the $+x$ direction when $\Delta T>0$. The thermal Hall signal should thus be negative in sign. As is shown in Appendix D, the thermal Hall signal contributed by the fluctuation of vector spin chirality is also negative in sign. 

We then assess the magnitude of the thermal Hall signal. For this purpose, let us introduce the dimensionless function $\gamma_{s}(t)$ as a measure of the strength of the fluctuation in the scalar spin chirality in the system. $\gamma_{s}(t)$ is defined as
\begin{equation}
\gamma_{s}(t)=\frac{1}{4NZ}\sum_{n,m}\frac{e^{- \frac{\epsilon_{n}}{t}}-e^{- \frac{\epsilon_{m}}{t}}}{\epsilon_{m}-\epsilon_{n}}|\langle n | \hat{\mathcal{C}} |m \rangle |^{2},
\end{equation}
in which $\epsilon_{n}=\frac{E_{n}}{J}$ and $t=\frac{k_{B}T}{J}$ are the eigenvalues and thermal energy measured in unit of $J$. We then have
 $\frac{|\kappa_{xy}|}{T}=\frac{J_{3}\sin\Phi_{3}}{J}\frac{k_{B}^{2}}{\hbar}f_{s}(t)$, in which $f_{s}(t)=-\frac{1}{2t}\frac{\partial \gamma_{s}(t)}{\partial t}$. One find that the contribution to $\kappa_{xy}$ from the vector spin chirality fluctuation can be written in exactly the same form(See Appendix D), with $\gamma_{s}(t)$ replaced by $\gamma_{v}(t)$ defined by
 \begin{eqnarray}
\gamma_{v}(t)&=&\frac{1}{Z}\sum_{n} e^{-\frac{\epsilon_{n}}{t}} \langle n | (\mathrm{S}_{i}\times\mathrm{S}_{j}) \cdot \ [ \ (\mathrm{S}_{n}\times\mathrm{S}_{p})\\
&+&(\mathrm{S}_{m}\times\mathrm{S}_{n})+(\mathrm{S}_{m}\times\mathrm{S}_{p})+(\mathrm{S}_{t}\times\mathrm{S}_{n})\ ]\ |n\ \rangle,\nonumber
\end{eqnarray}
in which the meaning of the site indices can be found from Fig. 1. Thus the magnitude of the thermal Hall conductivity is given by
 \begin{equation}
 \frac{|\kappa_{xy}|}{T}=\frac{J_{3}\sin\Phi_{3}}{J}\frac{k_{B}^{2}}{\hbar}\times(-\frac{1}{2t})\frac{\partial \gamma(t)}{\partial t}.
\end{equation}
Here $\gamma(t)=\gamma_{s}(t)+\gamma_{v}(t)$. A nonzero value of
$\frac{|\kappa_{xy}|}{T}$ in the low temperature limit should thus imply that $\gamma(t)\approx \gamma(0)-f(0)t^{2}$ in the same limit, a behavior naturally expected if the excitation spectrum of the system has a linear density of state at low energy. To be more quantitative, $\gamma(t)$ can be estimated from the measured thermal Hall signal as
 \begin{equation}
 \gamma(0)-\gamma(t_{0})=\frac{2J}{J_{3}\sin\Phi_{3}}\frac{\hbar}{k_{B}^{2}}\int_{0}^{t_{0}} \frac{|\kappa_{xy}|}{T}\  t \ dt.
\end{equation}
At B = 15T, we have $J_{3} \sin \Phi_{3}\approx10^{-3}J$. If we assume that $J=130 meV$ and that $\gamma(t)$ is essentially zero around T=100 K, then from the integration of the experimental data of $\frac{|\kappa_{xy}|}{T}$ one find that $\gamma(0)\approx 0.6$, which is a rather reasonable number(See Appendix E). In fact, from exact diagonalization calculation on a $4\times4$ cluster, we get $\gamma_{s}(0)=0.24$ and $\gamma_{v}(0)\approx 0.16$ . We note that the $4\times4$ cluster has a rather large finite size gap of about $0.6J$. With the increase of the cluster size, the energy scale of spin chirality fluctuation will decrease and their correlation length in space will increase, we thus expect a large value of $\gamma(0)$ in the thermodynamic limit.

When the bulk spin excitation of the system is gapped, a nonzero $\frac{\kappa_{xy}}{T}$ in the zero temperature limit would imply the existence of gapless chiral mode on the edge. For example, in the case of La$_{2}$CuO$_{4}$, there is bulk spin gap of $26$ K, while no sign of saturation in the increase of $\frac{|\kappa_{xy}|}{T}$ is found even around 10 K. This strongly suggests the existence of a gapless chiral edge mode. However, we note that the temperature dependence of $\frac{|\kappa_{xy}|}{T}$ varies between different families of cuprates. For example, in the case of Nd$_{2}$CuO$_{4}$, $\frac{|\kappa_{xy}|}{T}$ is found to approach zero in the zero temperature limit after rising to a peak at a temperature of the order of 10 K\ \cite{SNS}. The existence of gapless chiral edge mode is thus not universal. We will leave the analysis of such an edge physics to future works. In Appendix F, we present a possible effective model for such gapless chiral edge mode.

In conclusion, the giant negative thermal Hall signal observed in the pseudogap phase of the high-T$_{c}$ cuprates can be naturally understood as the orbital magnetic response of the local spin system. According this study, the giant thermal Hall signal is nothing but a measure of the spin chirality fluctuation in the system. There are many lessons to be learned from such an understanding. Firstly, the universal observation of the giant negative thermal Hall signal in and only in the  pseudogap phase implies that the antiferromagnetic correlation between the local spin is at the root of the pseudogap phenomena. Secondly, the tradeoff between the Hall signal and the thermal Hall signal with decreasing doping in the pseudogap phase implies that the transmutation between the local spin and the itinerant quasiparticle character of the electron at low energy is the key to understand the exotic behavior of the pseudogap phase. Thirdly, it implies that the pseudogap phenomena in cuprates is always accompanied by strong spin chirality fluctuation, which is suppressed only at sufficiently large doping when the local spin character of the electron is totally replaced by the itinerant quasiparticle character in the low energy physics. Lastly, we note that the effect discussed in this work is beyond the description of either the linear spin wave theory or the RVB mean field theory. The well known no-go theorem of thermal Hall effect on edge-sharing lattice is thus just a theoretical artifact of such approximations.

We acknowledge the support from the grant NSFC 11674391, the Research Funds of Renmin University of China, and the grant National Basic Research Project
2016YFA0300504. We would also like to thank Masataka Kawano, Chisa Hotta, Ga\"el Grissonnanchefor, Yuan-Ming Lu, Dung-Hai Lee, Patrick Lee, Subir Sachdev, Fu-Chun Zhang, Guang-Ming Zhang, Yi-Feng Yang, Long Zhang, Shou-Shu Gong, Bo Gu and Sadamichi Maekawa for useful comments. We are especially grateful to KITS at UCAS for organizing the discussion of the giant thermal Hall effect.

\section*{APPENDIXES}

\subsection{The classical ground state of $H_{J}+H_{3}$}
The Hamiltonian $H_{J}+H_{3}$ can be written as the sum of terms contributed by each triangle 
\begin{equation}
H=\sum_{\Delta_{i,j,k}}\ \mathrm{S}_{i}\cdot[\frac{J}{4}\ (\mathrm{S}_{j}+\mathrm{S}_{k})+J_{3} \sin\Phi_{3}\ (\mathrm{S}_{j}\times\mathrm{S}_{k})],
\end{equation}
in which site $j,k$ are nearest neighbors of site $i$. The factor $4$ comes from the fact that each bond of the square lattice is shared by four triangles. If we treat the spin as classical vectors of length $\frac{1}{2}$, we can easily prove that the classical ground state of $H$ is unchanged by $H_{3}$ even for rather large value of $J_{3}\sin\Phi_{3}$. In fact, it can be shown that for $J_{3}\sin\Phi_{3}<J/2$, the lowest energy state of each triangle is perfectly Neel ordered, so that the global minimum of the system must also be perfectly Neel ordered. 

From numerical optimization, we find that the critical value of $J_{3}\sin\Phi_{3}$ to have a nonzero spin chirality is $2J$, which is much larger than the above exact lower bound. For $J_{3}\sin\Phi_{3}>2J$ , the classical ground state has the form illustrated in Fig. 3. The spins on nearest neighboring sites are always orthogonal to each other in this phase. The distribution of spin chirality is not uniform in this phase. More specifically, in each plaquette of the square lattice, the spin chirality in the two triangles sharing the red bond reaches its largest possible value of $\frac{1}{8}$, while the spin chirality in the two triangles sharing the blue bond is zero.

\begin{figure}
\includegraphics[width=9cm]{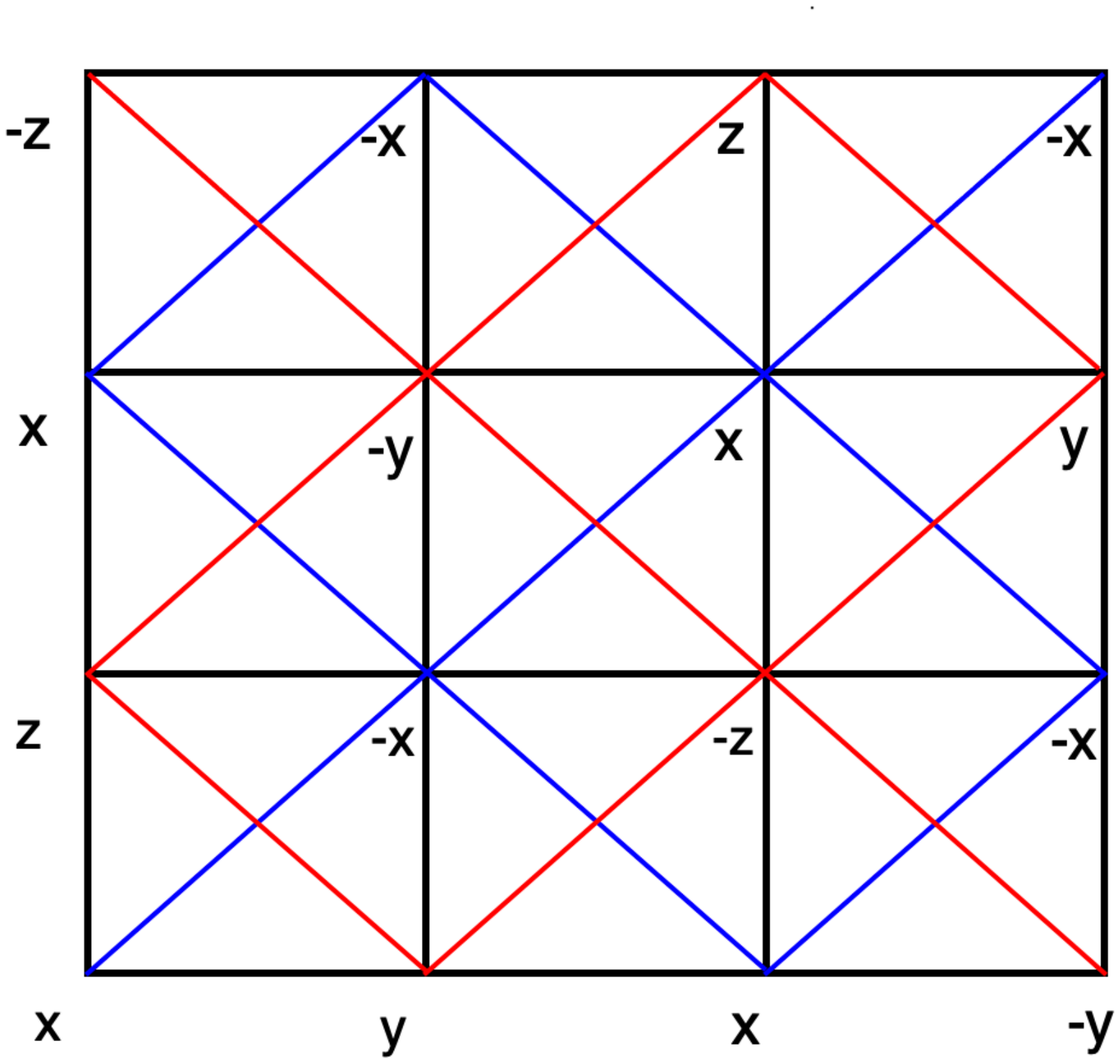}
\caption{The classical ground state of $H_{J}+H_{3}$ for $J_{3}\sin\Phi_{3}>2J$. The spin direction on each site is indicated explicitly. Spins on nearest neighboring sites are always orthogonal to each other in this phase. In each plaquette of the square lattice, the spin chirality in the two triangles sharing the red bond reaches its largest possible value of $\frac{1}{8}$, while the spin chirality in the two triangles sharing the blue bond is zero. }
\end{figure}

\subsection{Cancellation of the mean field contribution from $H_{3}$ in the Schwinger Boson mean field theory}
A representation of $H_{3}$ involving both the pairing field $\hat{A}_{i,j}$ and the hopping field $\hat{B}_{i,j}$ is given by
\begin{eqnarray}
\mathrm{S}_{i}\cdot (\mathrm{S}_{j}\times\mathrm{S}_{k})=\frac{1}{3\sqrt{2}i} \{\hat{A}^{\dagger}_{k,i}\hat{A}_{i,j}\hat{B}_{j,k}
+\hat{A}^{\dagger}_{i,j}\hat{A}_{j,k}\hat{B}_{k,i}\nonumber\\
+\hat{A}^{\dagger}_{j,k}\hat{A}_{k,i}\hat{B}_{i,j}-h.c.\}.
\end{eqnarray}  
In such a representation, it seems promising that $H_{3}$ can generate a linear-in-field thermal Hall signal. More specifically, the expectation value of $\hat{A}^{\dagger}_{i,j}$, $\hat{A}_{j,k}$ and $\hat{B}_{k,i}$ are all nonzero in the state described by ansatz Eq. (3)(see Fig. 4). We can thus modify their phases to accommodate a nonzero gauge flux $\Phi_{i,j,k}=\arg (A^{*}_{i,j}A_{j,k}B_{k,i})$. However, one find that no matter how we adjust the phases of the six fields  $A_{i,j}, A_{j,k}, A_{l,k}, A_{il}, B_{j,k}$ and $B_{i,l}$, we always have $\Phi_{i,j,k}+\Phi_{j,k,l}+\Phi_{k,l,i}+\Phi_{l,i,j}=0\mod(2\pi)$. The coupling of the magnetic field to the spin chirality still vanishes at the linear order in the mean field approximation.
\begin{figure}
\includegraphics[width=6cm]{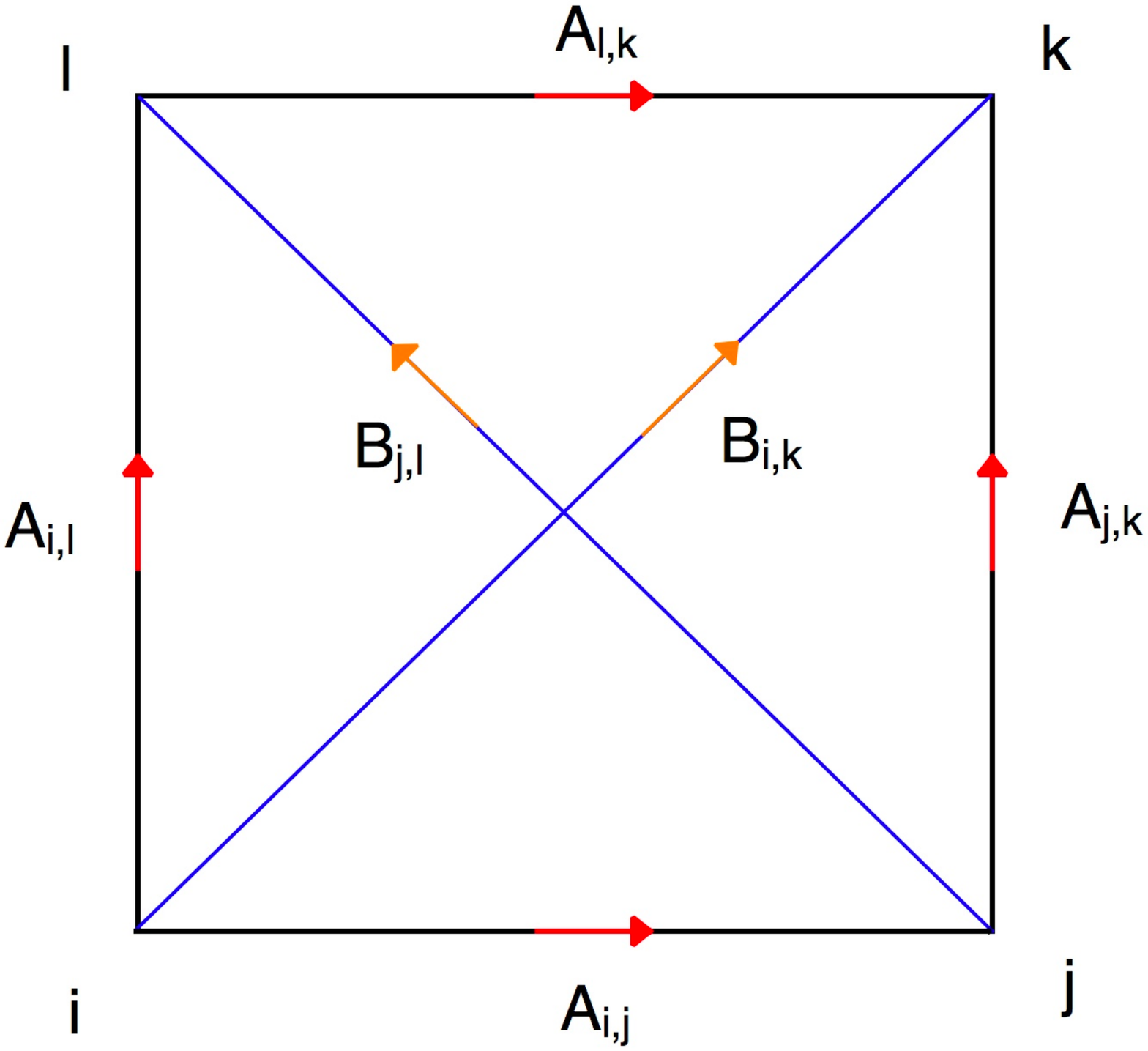}
\caption{Cancellation of the mean field contributions from $H_{3}$ around the mean field ansatz Eq. (3). In the mean field ground state described by such a ansatz, $B_{i,j}=0$ between nearest neighboring site and $A_{i,j}=0$ between next nearest neighboring sites. The phases of the pairing and hopping fields can be modified by the magnetic field. The gauge flux enclosed in a triangle(say, $\bigtriangleup_{i,j,k}$) is defined as $\Phi_{i,j,k}=\arg(A^{*}_{i,j}A_{j,k}B_{k,i})$. It is then straightforward to show that no matter how we adjust the phases of the pairing and the hopping fields, we always have $\Phi_{i,j,k}+\Phi_{j,k,l}+\Phi_{k,l,i}+\Phi_{l,i,j}=0\mod(2\pi)$.}
\end{figure}

\subsection{The Schwinger Boson mean field ground state of $H_{J}+H_{3}$}
To check if a modified mean field ansatz can accommodate a nonzero spin chirality at weak field, we have performed unrestricted mean field search of Schwinger Boson mean field ansatz for $H_{J}+H_{3}$ on a finite cluster. Both the phase and the amplitude of the nearest and the next-nearest RVB parameters are allowed to vary. It is found that the expectation value of $H_{3}$ is still identically zero for $J_{3}\sin\Phi_{3}<J_{c}$, where $J_{c}\approx 1.2J$(see Fig. 5). When $J_{3}\sin\Phi_{3}>J_{c}$, the antiferromagnetic correlation between nearest neighboring spins are strongly reduced.  At the same time, the distribution of the spin chirality in the triangles is found to exhibit the same pattern as that shown in Fig. 3. More specifically, in each plaquette of the lattice, the two triangles sharing a red bond have a much larger spin chirality than that of the two triangles sharing a blue bond. The Schwinger Boson mean field phase diagram of $H_{J}+H_{3}$ is thus qualitatively equivalent to that of the classical model.

\begin{figure}
\includegraphics[width=8cm]{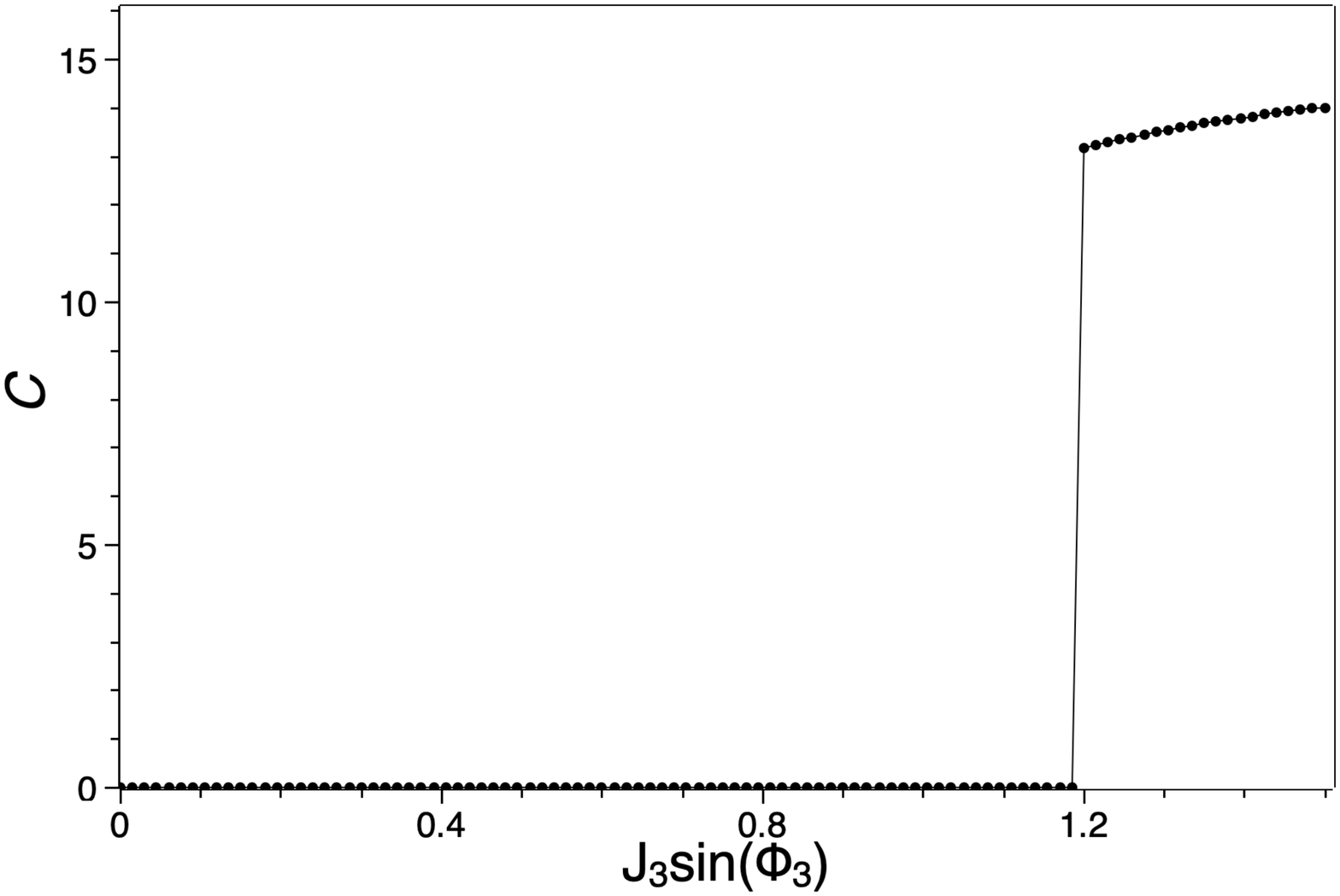}
\caption{The spin chirality as a function of $J_{3}\sin\Phi_{3}$ as calculated from numerical optimization of Schwinger Boson mean field energy with arbitrary nearest and next-nearest RVB parameters on a $4\times4$ cluster. Here we have set $J=1$ as the unit of energy.}
\end{figure}

\subsection{The energy current operator at the first order of the applied magnetic field}
In the main text, we have presented the energy current operator contributed by the Heisenberg part of the model. The $H_{3}$ term can also contribute to the energy current at the linear order. We find that such a contribution is given by
\begin{eqnarray}
J^{E,(1)}_{i,j}&=&-\frac{JJ_{3}\sin\Phi_{3}}{4}(\mathrm{S}_{i}\times\mathrm{S}_{j})\cdot\{ \nonumber\\
&&(\mathrm{S}_{i}+\mathrm{S}_{j})\times[(\mathrm{S}_{p}+\mathrm{S}_{l})-(\mathrm{S}_{k}+\mathrm{S}_{n}))]\nonumber\\
&+&(\mathrm{S}_{q}-\mathrm{S}_{m})\times[(\mathrm{S}_{k}+\mathrm{S}_{l})-(\mathrm{S}_{p}+\mathrm{S}_{n}))]\nonumber\\
&+&[(\mathrm{S}_{t}\times\mathrm{S}_{n})-(\mathrm{S}_{r}\times\mathrm{S}_{l})]+[(\mathrm{S}_{p}\times\mathrm{S}_{u})-(\mathrm{S}_{k}\times\mathrm{S}_{s})]\nonumber\\
&+&2[(\mathrm{S}_{n}\times\mathrm{S}_{p})-(\mathrm{S}_{l}\times\mathrm{S}_{k})]\ \  \},
\end{eqnarray}
in which the meaning of the site indices can be found from Fig. 1. $J^{E,(1)}_{i,j}$ has been written in a form which is explicitly odd under the reflection with respect to a line connecting site $i$ and $j$. Thus in the bulk of the system we have $\langle J^{E,(1)}_{i,j}\rangle=0$. On the upper edge of the system, the energy current operator becomes
\begin{eqnarray}
J^{E,(1)}_{i,j}&=&-\frac{JJ_{3}\sin\Phi_{3}}{4}(\mathrm{S}_{i}\times\mathrm{S}_{j})\cdot\{ \nonumber\\
&&(\mathrm{S}_{i}+\mathrm{S}_{j})\times(\mathrm{S}_{p}-\mathrm{S}_{n})\nonumber\\
&+&(\mathrm{S}_{m}-\mathrm{S}_{q})\times(\mathrm{S}_{p}+\mathrm{S}_{n})\nonumber\\
&+&(\mathrm{S}_{t}\times\mathrm{S}_{n})+(\mathrm{S}_{p}\times\mathrm{S}_{u})\nonumber\\
&+&2(\mathrm{S}_{n}\times\mathrm{S}_{p})\ \  \}.
\end{eqnarray}
At the leading order in $\Phi_{3}$, we can approximate $\langle J^{E,(1)}_{i,j}\rangle$ with $\langle J^{E,(1)}_{i,j}\rangle_{\mathrm{HAF}}$ - the expectation value of $ J^{E,(1)}_{i,j}$ in ground state of $H_{J}$. Using time reversal symmetry and mirror symmetry about the y-axis, $\langle J^{E,(1)}_{i,j}\rangle_{\mathrm{HAF}}$ can be simplified to the form of
\begin{eqnarray}
&\langle J^{E,(1)}_{i,j}\rangle_{\mathrm{HAF}}=-\frac{JJ_{3}\sin\Phi_{3}}{2}\langle \ (\mathrm{S}_{i}\times\mathrm{S}_{j})\cdot\\
& [ (\mathrm{S}_{n}\times\mathrm{S}_{p})+(\mathrm{S}_{m}\times\mathrm{S}_{n})+(\mathrm{S}_{m}\times\mathrm{S}_{p})+(\mathrm{S}_{t}\times\mathrm{S}_{n})]\ \rangle_{\mathrm{HAF}}.\nonumber
\end{eqnarray}
This is just the correlation of vector spin chirality on different bonds. $\langle J^{E,(1)}_{i,j}\rangle$ thus measures the fluctuation of the vector spin chirality in the system. 

Among the four terms in Eq. (14), $\langle (\mathrm{S}_{i}\times\mathrm{S}_{j})\cdot(\mathrm{S}_{n}\times\mathrm{S}_{p}) \rangle_{\mathrm{HAF}}$ is obviously the largest. In a background with dominate antiferromagnetic correlation, the expectation value of $\langle (\mathrm{S}_{i}\times\mathrm{S}_{j})\cdot(\mathrm{S}_{n}\times\mathrm{S}_{p}) \rangle_{\mathrm{HAF}}$ must be positive. We thus conclude that the energy current contributed by $J^{E,(1)}_{i,j}$ is also running in the $-x$ direction on the upper edge of the system. The thermal Hall signal contributed by the fluctuation of the vector spin chirality is thus also negative in sign.  

\begin{figure}
\includegraphics[width=8.5cm]{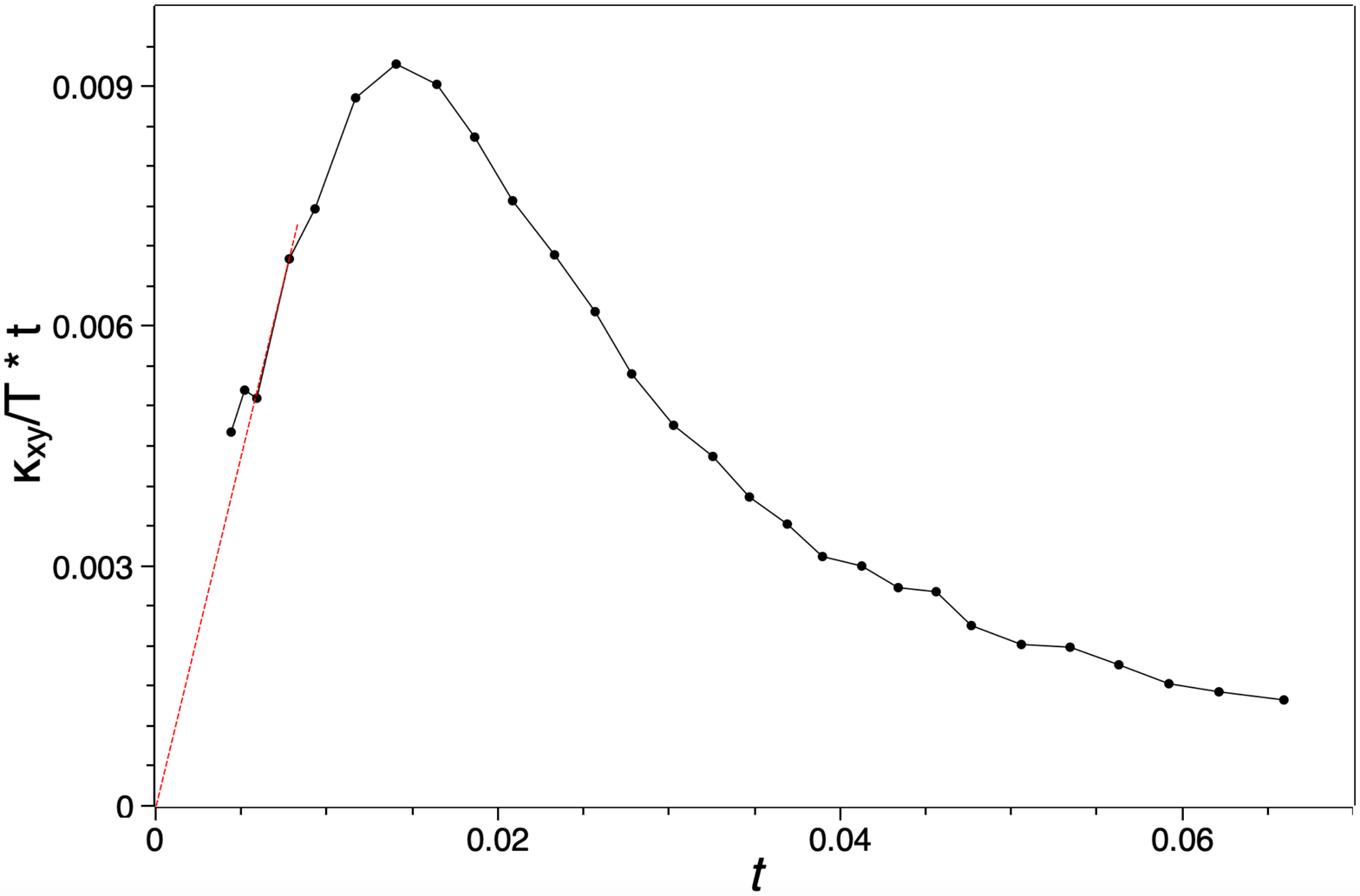}
\includegraphics[width=8.5cm]{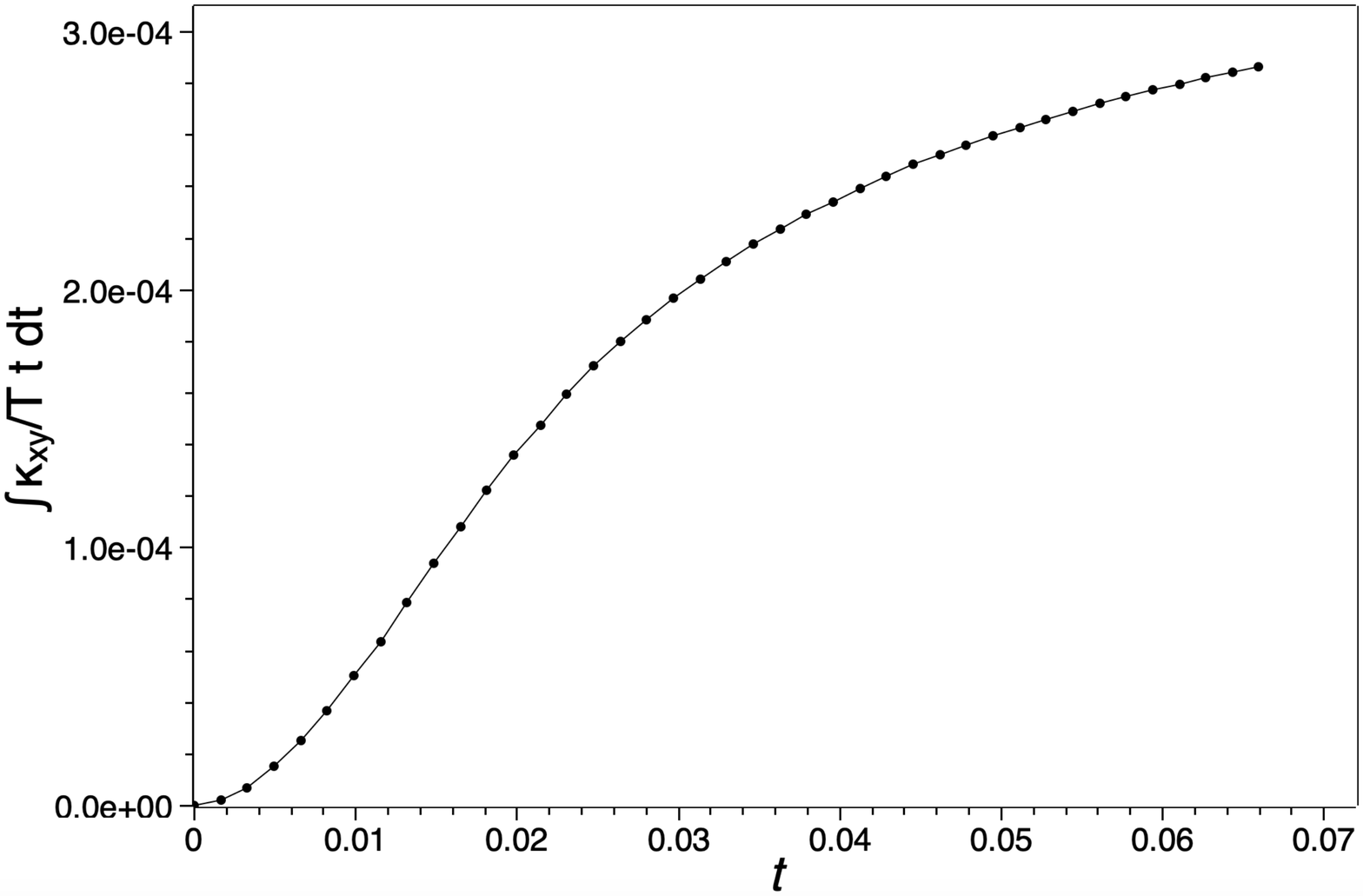}
\caption{Upper panel: The product of the observed thermal Hall signal $\frac{\kappa_{xy}}{T}$(in unit of $\frac{k_{B}^{2}}{\hbar}$) with the reduced temperature $t$(Here assume $J=130 \ meV$). The experimental data is taken from [\onlinecite{Taillefer}] and measured at B=15T on La$_{2}$CuO$_{4}$. The red dashed line is a linear extrapolation in the low temperature limit adopted here to perform the numerical integration. Lower panel: the integration of the thermal Hall conductance over temperature $\int\frac{\kappa_{xy}}{T} \ t \ dt$ from the experimental data. }
\end{figure}

\subsection{Estimation of the strength of spin chirality fluctuation from the observed thermal Hall signal}
From the main text we know that the strength of the spin chirality fluctuation in the system can be estimated from the measured thermal Hall signal. More specifically, we have
 \begin{equation}
 \gamma(0)-\gamma(t_{0})=\frac{2J}{J_{3}\sin\Phi_{3}}\frac{\hbar}{k_{B}^{2}}\int_{0}^{t_{0}} \frac{|\kappa_{xy}|}{T}\  t \ dt.
\end{equation}
In Fig. 6, we have reproduced the experimental data measured at B=15T on La$_{2}$CuO$_{4}$ from [\onlinecite{Taillefer}]. The thermal Hall conductivity is measured in unit of $\frac{k_{B}^{2}}{\hbar}$ and the temperature is measured in unit of $J$, with $J=130 \ meV$. We have adopted a linear extrapolation for $\frac{\kappa_{xy}}{T}\  t$ at low temperature to perform the numerical integration, the result of which is shown in the lower panel. If we assume that the spin chirality fluctuation is significantly disrupted at T=100K, then from Fig. 6 and the fact $J_{3}\sin \Phi_{3}\approx 10^{-3} J$ at B=15T, we can estimate that $\gamma(0)\approx 0.6$, a reasonable value consistent with the result from exact diagonalization calculation on the Heisenberg model. 

\begin{figure}
\includegraphics[width=8.5cm]{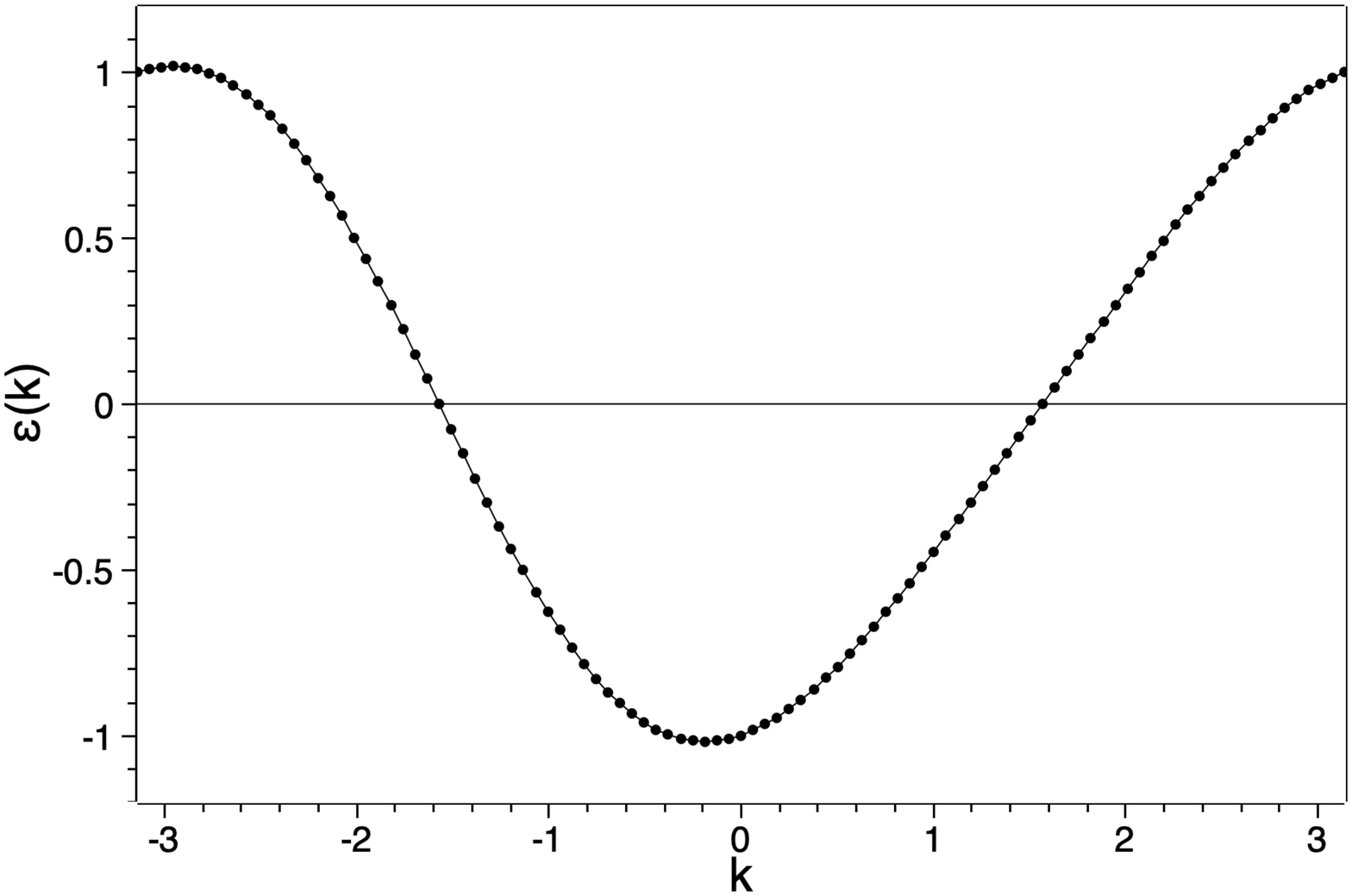}
\caption{Illustration of the mean field dispersion $\epsilon(k)$. Here we set $J_{1}u=1$ as the unit of energy. The dispersion curve shown here is for $J_{c}u^{2}=0.2$. Note that while reflection symmetry is broken, the particle-hole symmetry still holds.}
\end{figure}

\subsection{An effective model for the gapless chiral edge mode}
We model the edge degree of freedom with a $s=\frac{1}{2}$ spin chain. The influence of the nonzero spin chirality in the bulk can be modeled by a ring-exchange term on the chain. The model Hamiltonian reads
\begin{equation}
H=J_{1}\sum_{i}\mathrm{S}_{i}\cdot\mathrm{S}_{i+1}+J_{c}\sum_{i}\mathrm{S}_{i}\cdot(\mathrm{S}_{i-1}\times\mathrm{S}_{i+1}).
\end{equation}
We note that both $J_{1}$ and $J_{c}$ should be understood as phenomenological parameters, rather than the bare exchange coupling constant in the original 2D model. However, we expect $J_{c}$ to be linear in $\Phi_{3}$ at weak field.

It is well known that the Fermionic RVB theory works extremely well for the spin-$\frac{1}{2}$ antiferromagnetic Heisenberg chain. It is thus reliable to treat the ring-exchange term at weak field as a perturbation around this saddle point. More specifically, we introduce Fermionic spinon operator $f_{i,\alpha}$ to represent the spin operator as $\mathrm{S}_{i}=\frac{1}{2}\sum_{\alpha,\beta}f^{\dagger}_{i,\alpha}\sigma_{\alpha,\beta}f_{i,\beta}$. The Hamiltonian written in terms of the spinon operator has the form of
\begin{eqnarray}
H&=&-\frac{J_{1}}{2}\sum_{i}\hat{u}^{\dagger}_{i,i+1}\hat{u}_{i,i+1}\nonumber\\
&+&\frac{J_{c}}{4i}\sum_{i}(\hat{u}_{i,i+1}\hat{u}_{i+1,i-1}\hat{u}_{i-1,i}-h.c.).
\end{eqnarray}
Here $\hat{u}_{i,j}=\sum_{\alpha}f^{\dagger}_{i,\alpha}f_{j,\alpha}$.
The Fermionic RVB saddle point of the spin-$\frac{1}{2}$ antiferromagnetic Heisenberg chain is a simple 1D Fermi sea dictated by the RVB parameter $u=\langle \hat{u}_{i,i+1}\rangle$. Thus to the lowest order in $J_{c}$, the mean field Hamiltonian is given by
\begin{eqnarray}
H_{\mathrm{MF}}&=&-\frac{J_{1}u}{2}\sum_{i}(\hat{u}_{i,i+1}+h.c.)\nonumber\\
&+&\frac{J_{c}u^{2}}{4i}\sum_{i}(\hat{u}_{i+1,i-1}-h.c.).
\end{eqnarray}
Here we have used the fact that $\langle \hat{u}_{i-1,i+1}\rangle=0$ in the saddle point of the pure Heisenberg chain. The mean field dispersion of $H_{\mathrm{MF}}$ is given by $\epsilon(k)=-J_{1}u\cos k+\frac{J_{c}u^{2}}{2}\sin 2k$ and is plotted in Fig. 7. The Fermi velocity on the two Fermi points are different. The model thus has a gapless chiral spin liquid ground state.


\begin{thebibliography}{99}
\bibitem{phonon}The observed thermal Hall signal can not be attributed to the skew scattering of phonon, since the change in the longitudinal thermal conductivity in the magnetic field is found to be negligibly small\cite{Taillefer}.
\bibitem{DM}The Zeeman coupling and the anisotropic DM interaction are neglected in our discussion, since previous studies show that these couplings can not account for the observed thermal Hall effect\cite{PALee,Sachdev}.
\bibitem{Fermion}Here we adopt a Bosonic, rather than a Fermionic RVB picture. In the Fermionic RVB picture, the Fermionic spinon would unavoidably experience a large SDW gap in the presence of the static antiferromagnetic order, which would prohibit the system to exhibit a linear-in-field thermal Hall signal at low temperature\cite{Xu}.
\bibitem{firstorder}$H_{3}$ can generate an additional contribution to the energy current at the linear order of the external field. Such a contribution is proportional to the strength of vector spin chirality fluctuation in the system. The corresponding thermal Hall signal is also negative. A discussion on this issue is presented in Appendix D.
\bibitem{relax}The expectation value of the spin chirality operator near the boundary of the system will be relaxed and thus be different from that in the bulk. However, such a relaxation effect will not change our conclusion in a qualitative way.
\bibitem{leq}Here we have neglected the contribution to the thermal Hall signal from the usual Kubo term of the magnon system. This is reasonable since the static ordering pattern that the magnon experience is still non-chiral in the magnetic field, even if the system has already established a nonzero spin chirality at the quantum level.
\end{thebibliography}
\end{document}